\documentclass[a4paper,10pt,aps,prd,twocolumn,preprintnumbers,nofootinbib,showkeys,superscriptaddress,longbibliography]{revtex4-1}

\pdfoutput=1
\usepackage[dvipsnames]{xcolor}
\usepackage{graphicx,epic,eepic,epsfig}
\usepackage[export]{adjustbox}
\usepackage{tikz,contour}
\usepackage{feynmf}
\usetikzlibrary{shapes,arrows,positioning,automata,backgrounds,calc,er,patterns}
\usepackage[compat=1.1.0]{tikz-feynman}
\usepackage{tikz}
\usepackage{amsmath}
\usepackage{fullpage}
\usepackage{relsize}
\usepackage[left=1.9cm,right=1.9cm,top=1.9cm,bottom=1.9cm]{geometry}
\usepackage{XCharter}
\usepackage[T1]{fontenc}
\usepackage{amsfonts,amsmath,amssymb,amsbsy,bm,mathtools,slashed,latexsym,esvect,simpler-wick}

\usepackage{float,multirow,array,cancel,dcolumn}
\usepackage{enumitem}
\usepackage{hyperref}
\hypersetup{colorlinks=true,citecolor=red,linkcolor=NavyBlue,urlcolor=NavyBlue}
\usepackage[caption=false,labelformat=simple]{subfig}

\usepackage{comment}
\usepackage{shorthand}
\usepackage{natbib}

\begin{document}
	
	\title{Multi-Higgs production from a theory of varying Higgs quartic and top Yukawa couplings}
	
	\author{Ulf Danielsson}
	\email{ulf.danielsson@physics.uu.se}
	\affiliation{Department of Physics and Astronomy, Uppsala University, Box 516, SE-751 20 Uppsala, Sweden}
	
	\author{Rikard Enberg}
	\email{rikard.enberg@physics.uu.se}
	\affiliation{Department of Physics and Astronomy, Uppsala University, Box 516, SE-751 20 Uppsala, Sweden}
	
	\author{Gunnar Ingelman}
	\email{gunnar.ingelman@physics.uu.se}
	\affiliation{Department of Physics and Astronomy, Uppsala University, Box 516, SE-751 20 Uppsala, Sweden}
	
	\author{Soumyadip Kundu}
	\email{soumyadip23@iisertvm.ac.in}
	\affiliation{Indian Institute of Science Education and Research Thiruvananthapuram, Vithura, Kerala, 695 551, India}
	
	\author{Tanumoy Mandal}
	\email{tanumoy@iisertvm.ac.in}
	\affiliation{Indian Institute of Science Education and Research Thiruvananthapuram, Vithura, Kerala, 695 551, India}
	
	\author{Subhadip Mitra}
	\email{subhadip.mitra@iiit.ac.in}
	\affiliation{Center for Computational Natural Sciences and Bioinformatics, International Institute of Information Technology, Hyderabad 500 032, India}
	\affiliation{Center for Quantum Science and Technology, International Institute of Information Technology, Hyderabad 500 032, India}
	
	\begin{abstract}
		\noindent 
		We study the collider phenomenology of a theory in which the Higgs quartic self-coupling and the top Yukawa coupling are \textit{varying couplings}, which arise dynamically from two scalar fields, $\phi_1$ and $\phi_2$. Depending on the scalar masses relative to twice the top-quark mass, the heavy scalars decay mainly to Higgs boson pairs ($hh$) or to top-quark pairs ($t\bar t$), giving two distinct signatures. We find sizeable cross sections for resonant $hh$ production, and also for the background-suppressed process with three Higgs bosons ($hhh$) from associated production, $pp \to h\phi_i \to hhh$. By recasting ATLAS resonance searches, we set bounds on the effective new-physics scales $\Lambda_1$ and $\Lambda_2$ at the TeV scale. We also show that the High-Luminosity LHC (HL-LHC) will probe this parameter space further, through combined measurements of the $hh$, $t\bar t$ and multi-Higgs channels. This gives a strong, multi-channel test of whether the electroweak sector has a dynamical origin.
	\end{abstract}
	\maketitle 
	\section{Introduction}\label{sec:introduction}
	\noindent
	For decades, physicists have speculated that dimensionless fundamental constants might vary across spacetime (see Refs.~\cite{Uzan:2002vq, Uzan:2010pm,Uzan:2024ded} for reviews). In a fundamental theory, one expects no free parameters---all coupling constants should emerge from the underlying dynamics. In string theory or quantum gravity, for instance, coupling constants are dynamical. The vacuum expectation values (vevs) of scalar fields determine all parameters of the low-energy effective theory. J.~Bekenstein first proposed a consistent framework for such varying-coupling theories~\cite{Bekenstein:1982eu,Bekenstein:2002wz}. He formulated a model with a spacetime-dependent fine-structure constant, where a real scalar field interacting with the electromagnetic sector drives the variation.
	
	These varying-coupling theories carry major cosmological implications. For example, varying Yukawa couplings might trigger a strong first-order electroweak phase transition---an essential ingredient for electroweak baryogenesis~\cite{Baldes:2016rqn, Braconi:2018gxo,Bruggisser:2018mus, Ellis:2019flb}. Similarly, a dynamical strong coupling could force QCD confinement near the phase transition, potentially driving baryogenesis~\cite{Ipek:2018lhm, Croon:2019ugf,Cline:2025bwe} and generating gravitational waves~\cite{Chatrchyan:2025wop,Gao:2024pqm,Sagunski:2023ynd}. As the early Universe cools, the scalar potential settles into a stable minimum. This traps the associated scalar field, freezing the coupling to the constant value we observe today.
	
	We previously explored a model with a varying electromagnetic coupling~\cite{Danielsson:2016nyy}, later generalising this framework to all Standard Model (SM) gauge couplings~\cite{Kimberly:2003rz,Shaw:2004hk,Danielsson:2019ftq,deLima:2026xmt}---a direction motivated by gauge coupling unification. Because the hierarchy problem hinges on the interplay between the Higgs boson and the top quark, we now shift our focus to these two sectors. Here, we explore the dynamic variation of the Higgs quartic coupling $\lambda_h$ and the top-quark Yukawa coupling $\lambda_t$. We recently introduced a minimal One Scalar Theory to describe this mechanism~\cite{Danielsson:2026ilg}. The dynamics of this scalar parallels that of dilaton models, where a light scalar emerges as the pseudo-Goldstone boson of spontaneously broken conformal invariance~\cite{Goldberger:2007zk, Barger:2011nu, Bellazzini:2012vz, Jung:2014zga, Giudice:2017fmj, Ahmed:2019csf}. This comprehensive paper expands upon that foundation, detailing the broader theoretical framework and the resulting massive scalar states in the particle spectrum.
	
	If these scalars are related to electroweak symmetry breaking (EWSB), we expect their masses to lie near the electroweak scale. This makes the scenario particularly interesting for both cosmology and colliders. We defer early-Universe dynamics to future work and focus here on collider phenomenology. Beyond the minimal single scalar driving both $\lambda_h$ and $\lambda_t$, we introduce a scenario with two distinct scalars. These scalars and their mixing generate signatures we can test at the LHC, using rare SM processes as sensitive probes. For a rare process, observing a rate that exceeds its tiny SM expectation provides a clear sign of physics beyond the SM (BSM). Multi-Higgs production offers a prime example: Because the SM cross section for di-Higgs production at the LHC is so small, any deviation in its rate or kinematics points directly to new interactions or particles. This makes di-Higgs searches a vital tool for probing BSM scenarios. Our framework also predicts potentially observable tri-Higgs signatures---events otherwise too suppressed in the SM to detect, even at the High-Luminosity LHC (HL-LHC).
	
	The remainder of this paper is organised as follows. In  Sec.~\ref {sec:model}, we introduce the theoretical framework for dynamically varying fundamental parameters. We detail our specific parameterisation of the Higgs quartic and top-quark Yukawa couplings, and formally define both the one-scalar (1ST) and two-scalar (2ST) models. In Sec.~\ref{sec:pheno}, we discuss the phenomenology of the new scalar states at the LHC, evaluating their primary production mechanisms and decay branching ratios (BRs). In Sec.~\ref {sec:exclu}, we confront our framework with the latest experimental data from the LHC, deriving current exclusion limits on the free parameters of the model. We then extend our analysis in Sec.~\ref {sec:reach} to estimate the projected sensitivity and discovery reach at the upcoming High-Luminosity LHC (HL-LHC). Finally, we summarise our findings and conclude in 
	Sec.~\ref {sec:summary}.
	
	\section{Varying $\lambda_h$ and $\lambda_t$, and other possibilities}
	\label{sec:model}
	\noindent
	In principle, any dimensionless coupling can exhibit spacetime dependence from an associated scalar field. In string theory, such fields are know as moduli fields. An effective field theory (EFT) framework containing massive scalar fields naturally captures this variation. In this paper we focus on the simultaneous variation of the Higgs quartic coupling, $\lambda_h$, and the top-quark Yukawa coupling, $\lambda_t$. We explore two scenarios. In the One Scalar Theory (1ST)~\cite{Danielsson:2026ilg}, a single real scalar field, $S_0$, governs both variations; in the Two Scalar Theory (2ST), the couplings vary independently, controlled by distinct real scalars that are allowed to mix. 
	
	We promote dimensionless couplings to spacetime-dependent functions, $\tilde{\lambda}_i = \lambda_i\epsilon_i(x)$ (no summation), where $\epsilon_i(x)$ is a dimensionless scalar field. The field $\epsilon_i$ takes a non-canonical, rescaling-invariant kinetic term of the form $\frac{1}{2}(\Lambda_i^2/\epsilon_i^2)(\partial_\mu \epsilon_i)^2$, a structure characteristic of moduli fields. By performing the field redefinition $\epsilon_i = \exp(S_i/\Lambda_i)$, we obtain a canonically normalised real scalar, $S_i$, which couples to the SM sector via dimension-five operators. Here, $\Lambda_i$ represents the characteristic energy scale of the new physics driving the variation.
	
	For a detailed description of the setup and field redefinitions used in this framework applied to the SM gauge couplings, see~\cite{Danielsson:2016nyy,Danielsson:2019ftq}. 
	
	\subsection{One scalar theory}
	\noindent
	While we explored the 1ST in Ref.~\cite{Danielsson:2026ilg}, we outline its core mechanics here to keep our discussion self-contained. Expanding the resulting interactions to leading order in $1/\Lambda_0$, the interaction Lagrangian is given by:
	\begin{align}
		\label{eq:1STLag}
		\mathcal{L} \supset&\ -\frac{S_0}{\Lambda_0} \left[ \lambda_h \left( v^2 h^2 + v h^3 + \frac{1}{4}h^4 \right) - \frac{\lambda_t}{\sqrt{2}}(v + h)\bar{t}t \right]\nonumber\\
		&\ + \mathcal{O}\left( \frac{1}{\Lambda_0^2} \right),
	\end{align}
	where $v$ is the Higgs vev. Since $S_0$ couples to the SM through these dimension-five operators, its production and decay modes get correlated, generating highly predictive and testable signatures at the LHC.
	
	The most general scalar potential containing $S_0$ and the SM Higgs doublet $\Phi$ takes the form,
	\begin{align}
		V(\Phi,S_0) &= \mu^2\lt(\Phi^\dagger\Phi\rt) + \lm\lt(\Phi^\dagger\Phi\rt)^2\nn\\
		&+ \sum_{n=1}^2 \lm_n S_0^n \lt(\Phi^\dagger\Phi\rt) + \sum_{n=1}^4 \lm^{\prime}_n S_0^n,
	\end{align}
	where $\mu^2$, $\lambda$, $\lambda_n$ and $\lambda_n^\prime$ are the free parameters of the scalar potential. After EWSB, the $\lambda_1 S_0 (\Phi^\dagger \Phi)$ interaction induces mixing between $S_0$ and the SM Higgs boson, $h$. Because current LHC data strongly constrain such mixing, we can ignore it and treat $S_0$ as a mass eigenstate. (Alternatively, one can also impose a discrete $\mathbb{Z}_2$ symmetry under which $S_0$ is odd to prevent it from acquiring a vev.)
	
	Ultimately, just two free parameters determine the entire collider phenomenology of the 1ST: the mass of the scalar $M_{0}$ and the single new physics scale $\Lambda_0$. This minimal parameter space allows us to test the model at the LHC through multi-Higgs and resonant di-top production, as detailed in Ref.~\cite{Danielsson:2026ilg}.
	
	\begin{table*}
		\caption{Effective couplings ($\kappa^i$) of the mass eigenstates $\phi_i$ to SM fields in the 2ST model. The parameters are derived from the Lagrangian terms specified in Eq.~\eqref{eq:2STlag}. We employ the shorthand notation $s_\alpha \equiv \sin\alpha$ and $c_\alpha \equiv \cos\alpha$. \label{tab:coupphi12}}
		\renewcommand{\arraystretch}{1.5}
		\begin{ruledtabular}
			\begin{tabular}{cccccc}
				Field & $\kappa^i_{hh}$ & $\kappa^i_{hhh}$ & $\kappa^i_{hhhh}$ & $\kappa^i_{t\bar{t}}$ & $\kappa^i_{ht\bar{t}}$ \\ \hline
				$\phi_1$ & $-\dfrac{\lambda_h v^2 c_\alpha}{\Lambda_1}$ & $-\dfrac{\lambda_h v c_\alpha}{\Lambda_1}$ & $-\dfrac{\lambda_h c_\alpha}{4\Lambda_1}$ & $-\dfrac{\lambda_t v s_\alpha}{\sqrt{2}\Lambda_2}$ & $-\dfrac{\lambda_t s_\alpha}{\sqrt{2}\Lambda_2}$ \\
				$\phi_2$ & $+\dfrac{\lambda_h v^2 s_\alpha}{\Lambda_1}$ & $+\dfrac{\lambda_h v s_\alpha}{\Lambda_1}$ & $+\dfrac{\lambda_h s_\alpha}{4\Lambda_1}$ & $-\dfrac{\lambda_t v c_\alpha}{\sqrt{2}\Lambda_2}$ & $-\dfrac{\lambda_t c_\alpha}{\sqrt{2}\Lambda_2}$ \\
			\end{tabular}
		\end{ruledtabular}
	\end{table*}
	
	\subsection{Two scalar theory}
	\noindent
	In the 2ST setup, the couplings $\lambda_h$ and $\lambda_t$ are dynamically generated by two distinct scalar fields: $\tilde{\lambda}_h = \lambda_h \epsilon_1$ and $\tilde{\lambda}_t = \lambda_t \epsilon_2$, where $\epsilon_i = \exp(S_i/\Lambda_i) \simeq 1 + S_i/\Lambda_i$ for $i \in \{1, 2\}$. Both $S_1$ and $S_2$ are canonically normalised scalar fields that do not acquire vevs. The fields $S_i$ serve as interaction eigenstates rather than mass eigenstates, as they may mix with each other and with the SM Higgs boson through the extended scalar potential. To satisfy the stringent constraints imposed by current LHC Higgs signal strength measurements, we assume that mixing between the $S_i$ fields and the Higgs is negligible \cite{ATLAS:2016neq,CMS:2018uag,ATLAS:2019slw}. We do, however, permit mixing between $S_1$ and $S_2$, which yields distinct collider signatures, as we discuss later. We define the physical scalar mass eigenstates, $\phi_1$ and $\phi_2$, via the orthogonal transformation:
	\begin{equation}\label{eq:mixmatrix}
		\begin{pmatrix}
			\phi_1 \\
			\phi_2
		\end{pmatrix} 
		=
		\begin{pmatrix}
			\cos\alpha & \sin\alpha \\
			-\sin\alpha & \cos\alpha
		\end{pmatrix}
		\begin{pmatrix}
			S_1 \\
			S_2
		\end{pmatrix},
	\end{equation}
	where the mixing angle $\alpha$ is determined by the parameters of the scalar potential. Expanding the fields to leading order in $1/\Lambda_i$, the interaction Lagrangian is given by:
	\begin{align}
		\label{eq:2STlag}
		\mathcal{L} \supset 
		&-\frac{S_1}{\Lambda_1}\left[\lambda_h\left(v^2 h^2 + v h^3 + \frac{1}{4}h^4\right)\right] \nonumber \\
		&-\frac{S_2}{\Lambda_2}\left[\frac{\lambda_t}{\sqrt{2}}(v+h)\bar{t}t\right] + \mathcal{O}\left(\frac{1}{\Lambda_i^2}\right).
	\end{align}
	Expressing the interaction eigenstates $S_i$ in terms of the physical mass eigenstates $\phi_i$ using Eq.~\eqref{eq:mixmatrix} yields the effective SM couplings of the physical scalars. These interactions are summarise in Table~\ref{tab:coupphi12}. Ultimately, the 2ST scenario features five free parameters: the scalar masses $M_1$ and $M_2$, the cutoff scales $\Lambda_1$ and $\Lambda_2$, and the mixing angle $\alpha$.
	
	\subsection{Other possibilities}
	\noindent
	In principle, all the dimensionless parameters in the SM can be set by a collection of multiple distinct scalars. However, to systematically study their phenomenology and focus on scenarios with potentially interesting collider signatures, we need some simplifying assumptions. Because our primary concern in this paper is multi-Higgs production, we focus on identifying the configuration that maximises the di-Higgs and tri-Higgs cross sections.
	
	Let us briefly consider possible frameworks with two varying parameters: $\lambda_h$ plus another dimensionless SM parameter, such as the three gauge couplings, or one of the five remaining quark Yukawa couplings. The scalar production cross section maximises when we vary the strong coupling. As a result, a priori, a scenario where a single scalar governs both the strong coupling and the Higgs quartic coupling seems promising. However, a large production cross section alone does not necessarily imply a large signal yield, which also depends on the relevant BRs. Among these eight possible pairings, varying $\lambda_h$ and $\lambda_t$ gives the largest signal yield in the region of our interest.
	
	\section{Production and decay}
	\label{sec:pheno}
	\noindent We use \textsc{FeynRules}~\cite{Alloul:2013bka} to implement the dimension-five effective interactions and interface the Universal FeynRules Output (UFO)~\cite{Degrande:2011ua} model files with the \texttt{MadGraph5\_aMC$@$NLO}~\cite{Alwall:2014hca} event generator. For the loop-induced processes, we perform the analytical integrations in \textsc{Package-X}~\cite{Patel:2015tea}. We estimate the cross sections in \textsc{Mathematica} with \texttt{NNPDF2.3}~\cite{Ball:2013hta,Hartland:2012ia} parton distribution functions (PDFs). Finally, for the higher-order QCD effects---particularly in the dominant gluon-fusion production modes---we calculate mass-dependent $K$-factors using \textsc{Higlu}~\cite{Spira:1995mt}. This procedure allows us to reliably scale our baseline cross sections to next-to-next-to-leading order (NNLO) accuracy as a function of the scalar mass $M_{i}$.
	\begin{figure*}[t]
		\centering
		\subfloat[]{\includegraphics[width=0.49\textwidth]{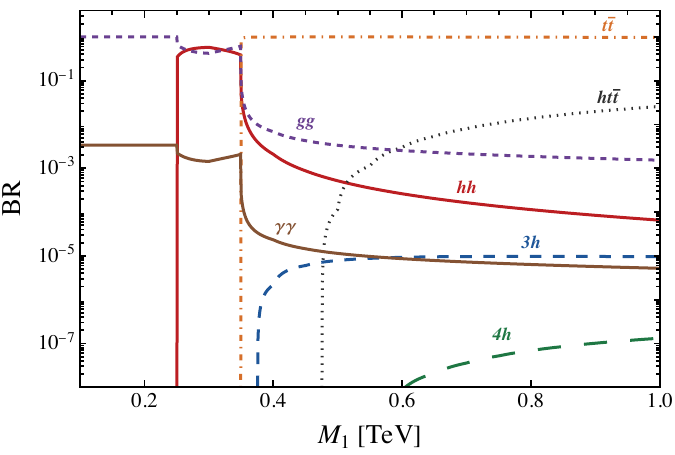}}\hfill
		\subfloat[]{\includegraphics[width=0.49\textwidth]{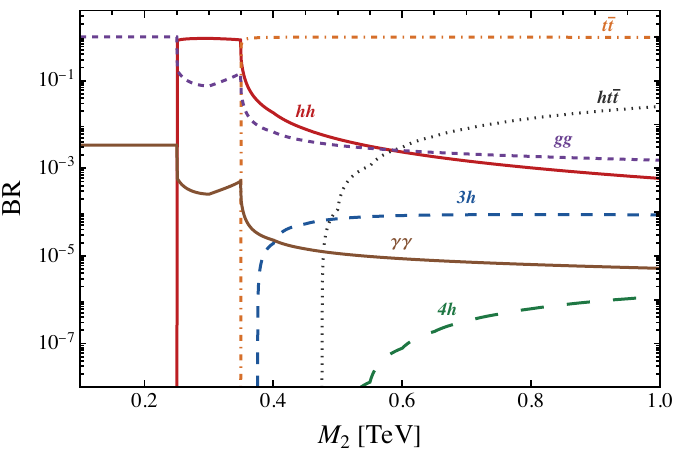}}\\
		\caption{\label{fig:2STmassBR}Branching ratios of the scalar $\phi_i$ as a function of $M_i$, $i\in \{1,2\}$ in the 2ST scenario ($c_\alpha=0.5,\Lambda_1 = 2$~TeV, $\Lm_2=1$~TeV). The scalar decays predominantly to $hh$ below the $t\bar{t}$ kinematic threshold, after which the $t\bar{t}$ mode rapidly dominates. }
		\medskip
		\subfloat[$\phi_1\rightarrow hh$  and $\phi_1\rightarrow t\bar{t}$]{\includegraphics[width=\columnwidth]{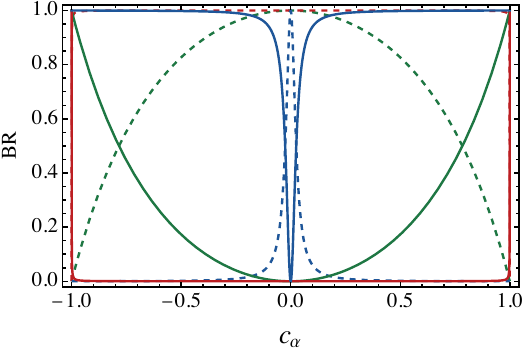}}\hspace{0.4cm}
		\subfloat[$\phi_2\rightarrow hh$  and $\phi_2\rightarrow t\bar{t}$]{\includegraphics[width=\columnwidth]{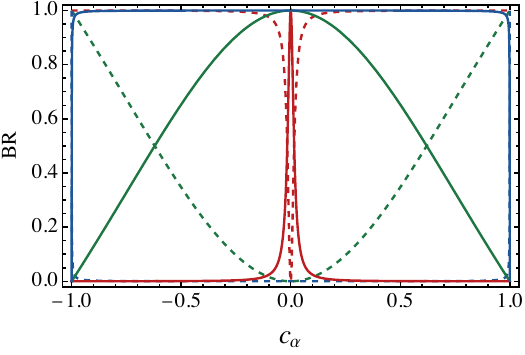}}\\
		\vspace{.3cm}
		\includegraphics[width=0.4\textwidth]{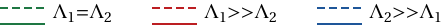}
		\caption{\label{fig:2STcalBR}Branching ratios of the 2ST mass eigenstates, $\phi_i$, into the $hh$ and $t\bar{t}$ channels as a function of $\cos\alpha$. The distributions are shown for $\Lm_1=\Lm_2$, $\Lambda_2 \gg \Lambda_1$ and $\Lambda_2 \ll \Lambda_1$. A fixed scalar mass of $M_{i} = 350$~GeV is assumed across all panels. Dashed lines represent the decay to $t\bar{t}$ whereas solid lines represent decay to $hh$.}
	\end{figure*}
	
	\subsection{Decay modes and branching ratios}
	\noindent 
	Depending on the mass $M_{i}$, the scalar can primarily decay into four two-body final states. For $M_{i} < 2m_h$, the loop-induced gluon-gluon ($gg$) and diphoton ($\gamma\gamma$) channels dominate. We estimate the partial widths as~\cite{Gunion:1989we}:
	\begin{align}
		\Gamma(\phi_i \to gg) &= \frac{\alpha_s^2 M_{i}^3}{32 \pi^3 \Lambda_2^2} \left| \tau \big[ 1 + (1-\tau) f(\tau) \big] \right|^2 \label{eq:GammaSgg}, \\
		\Gamma(\phi_i \to \gamma\gamma) &= \frac{\alpha^2 M_{i}^3}{256 \pi^3 \Lambda_2^2} \left| \frac{4}{3} F_t(\tau) \right|^2 \label{eq:GammaSaa},
	\end{align} 
	where $\tau = 4m_t^2/M_{i}^2$. Beyond the relevant kinematic thresholds, the tree-level di-Higgs ($hh$) and di-top ($t\bar{t}$) decay modes open up, producing the following partial widths:
	\begin{align}\label{eq:partial_widths}
		\Gamma(\phi_i \to hh) &= \frac{(\kappa_{hh}^i)^2}{8\pi M_{i}} \sqrt{1 - \frac{4m_h^2}{M_{i}^2}},  \\
		\Gamma(\phi_i \to t\bar{t}) &= \frac{3(\kappa_{tt}^i)^2}{8\pi} M_{i}\left(1 - \frac{4m_t^2}{M_{i}^2}\right)^{3/2},
	\end{align}
	where the effective, model-specific couplings $\kappa_{hh}^i$ and $\kappa_{tt}^i$ are shown in Table~\ref{tab:coupphi12}. Additionally, we estimate partial widths of higher-body decay modes such as $3h$, $h t\bar{t}$, and $4h$, which become kinematically accessible at higher masses, numerically using \texttt{MadGraph5\_aMC$@$NLO}. 
	
	The di-Higgs width is mass-suppressed ($\Gamma_{hh} \propto M_{i}^{-1}$), the di-top width scales linearly with mass ($\Gamma_{t\bar{t}} \propto M_{i}$) and $\Gamma_{gg}$ or $\Gamma_{\gamma\gamma}$ grows as $M_{i}^3$ due to the derivative nature of the dimension-five operator. The phenomenological impact of these scalings is visible in the BRs. Below the di-Higgs threshold ($M_{i} < 2m_h$), the scalar decays almost entirely into two gluons. In the intermediate window, $2m_h < M_{i} < 2m_t$, the scalar predominantly decays into $hh$. However, once the di-top threshold is crossed, the mass-linear scaling of $\Gamma_{t\bar{t}}$ causes the top-quark channel to rapidly overtake the $hh$ channel. At high masses, this kinematic advantage pushes the BR$(\phi_i \to t\bar{t})$ to nearly $100\%$, as illustrated in Fig.~\ref{fig:2STmassBR}. The minor contributions from the higher-body decay modes are also included in this figure for completeness. In Fig.~\ref{fig:2STcalBR}, we show the evolution of the BRs for the 2ST mass eigenstates into the $hh$ and $t\bar{t}$ final states as a function of the mixing parameter $\cos\alpha\equiv c_{\alpha}$. The dynamic interplay between these decay channels is highlighted across the three distinct phenomenological limits.
	
	\subsection{Production channels}
	\begin{figure*}[tb]
		\centering
		\subfloat[]{\includegraphics[width=0.42\textwidth]{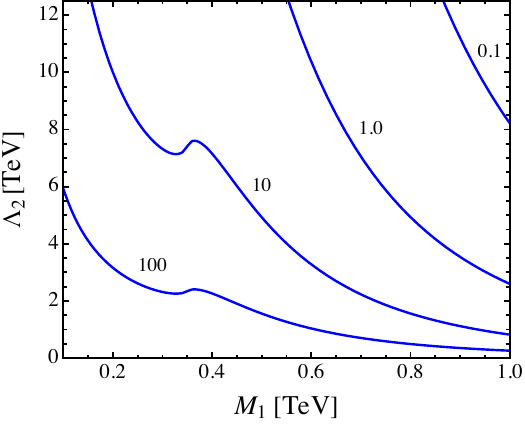}}
		\hspace{.6cm}
		\subfloat[]{\includegraphics[width=0.42\textwidth]{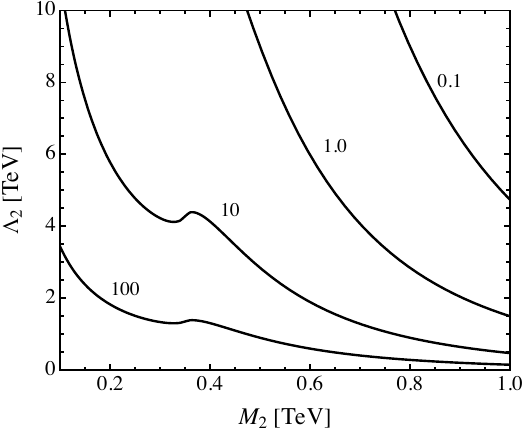}}
		\caption{Contours of the production cross section $\sigma(gg\to\phi_i)$ (in fb) in the $(M_i,\Lambda_2)$ plane for the 2ST model at the $\sqrt{s}=14$~TeV LHC, with $c_\alpha=0.5$.}
		\label{fig:contours}
		\centering
		\includegraphics[width=0.45\textwidth]{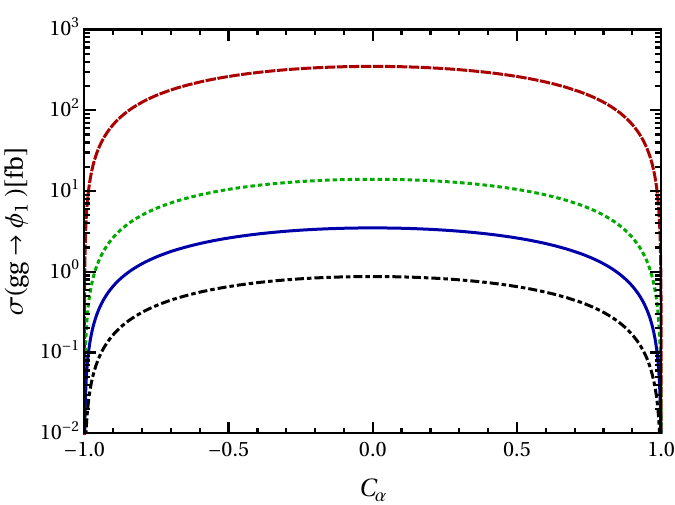}\hspace{1cm}
		\includegraphics[width=0.45\textwidth]{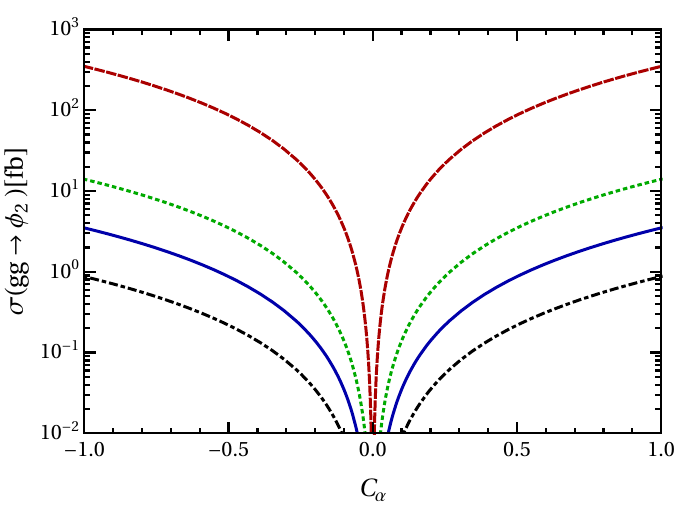}
		\label{fig:XSECALPHA}
		\vspace{0.3cm} 
		
		\includegraphics[width=0.55\textwidth]{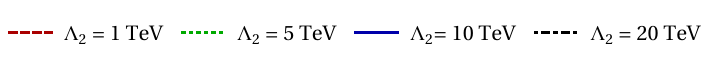}
		
		\caption{Total production cross sections (in fb) of the mass eigenstates $\phi_i$ in the 2ST model as functions of the mixing parameter $\cos\alpha$. The mass of the corresponding scalar state is fixed at $M_{i}=300$~GeV, and the cross sections are evaluated for the $\sqrt{s}=14$~TeV LHC.}
		\label{fig:CALxsec}
	\end{figure*}
	
	\noindent
	In the 2ST framework, the new scalar $\phi_i$ is primarily produced at the LHC via three mechanisms. Two of these are loop-induced gluon-gluon fusion (ggF) processes: inclusive single production ($gg\to\phi_i$) and associated production with a  SM Higgs boson ($gg \to h\phi_i$). The third mechanism is associated production with a top-quark pair ($pp \to t\bar{t}\phi_i$). For the inclusive single-resonance ggF channel, we calculate NNLO QCD cross section for $\sqrt{s} = 14$ TeV using \textsc{Higlu}, which provides the baseline production cross section for a scalar of mass $M_{i}$ assuming SM-like couplings. In our 2ST framework, however, the new scalar coupling to top quarks deviates from the SM top-Yukawa coupling. Therefore, we evaluate the baseline SM-like cross section over the mass range $M_{i} \in [100, 1000]$ GeV and obtain the cross section for our model by rescaling the \textsc{Higlu} output by the square of the particular coupling modifier from Table~\ref{tab:coupphi12}. 
	
	We calculate the relevant triangle and box loop diagrams to estimate the cross section of the associated $h\phi_i$ production. The required form factors are structurally identical to those in the SM di-Higgs production---see Appendix~\ref{sec:appendix} for the analytical expressions of amplitudes and form factors. We avoid the computationally expensive higher-order QCD calculation for this loop-induced process; instead, we use an approximate, mass-independent $K$-factor. Given the topological similarity between the $h\phi_i$ triangle diagram and inclusive single ggF production, we rescale the triangle contribution using the NNLO ggF $K$-factor evaluated at a representative mass of $M_{i} = 425$ GeV. We neglect higher-order QCD corrections to the box diagrams, as their contribution to the total cross section is minor. 
	
	For the associated $t\bar{t}\phi_i$ production channel, we estimate the baseline next-to-leading order (NLO) cross section for the Higgs Characterisation model~\cite{Demartin:2014fia} in \texttt{MadGraph5\_aMC$@$NLO}. We finally rescale the baseline result by the square of the corresponding $\phi_i t\bar{t}$ coupling modifier to obtain the required cross section.
	
	Since $\phi_i$ is produced either via top-quark loops (in the ggF channels) or in direct association with top quarks, the total production rate is highly sensitive to the $\phi_i t\bar{t}$ coupling, which is governed by $\Lambda_2$. Fig.~\ref{fig:contours} presents contour plots of $\sigma(gg\to\phi_i)$ in the $(M_{i}, \Lambda_2)$ plane, illustrating the specific parameter configurations required to yield constant production rates. In Fig.~\ref{fig:CALxsec}, we show the cross section of the ggF production of $\phi_i$ at a fixed scalar mass as a function of the mixing angle. The two mass eigenstates exhibit complementary behaviour governed by their respective trigonometric coupling scalings. Fig.~\ref{fig:CALxsec} also illustrates the cross section variation with $\cos\alpha$ across several benchmark values of $\Lambda_2$. 
	
	\begin{figure*}[tb]
		\centering
		\subfloat[]{\includegraphics[width=0.47\textwidth]{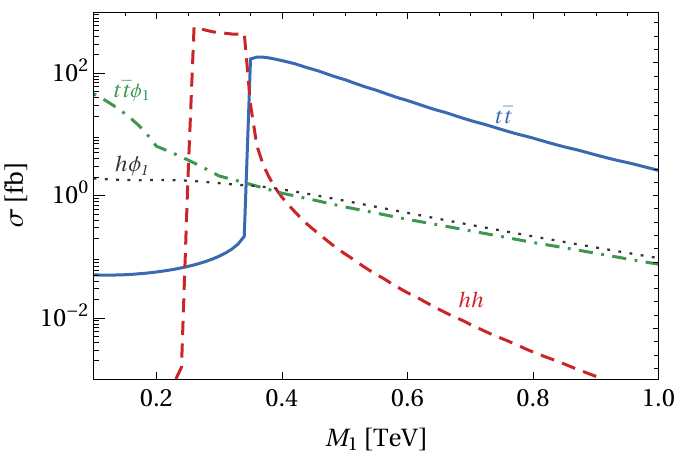}}\hspace{.5cm}
		\subfloat[]{\includegraphics[width=0.47\textwidth]{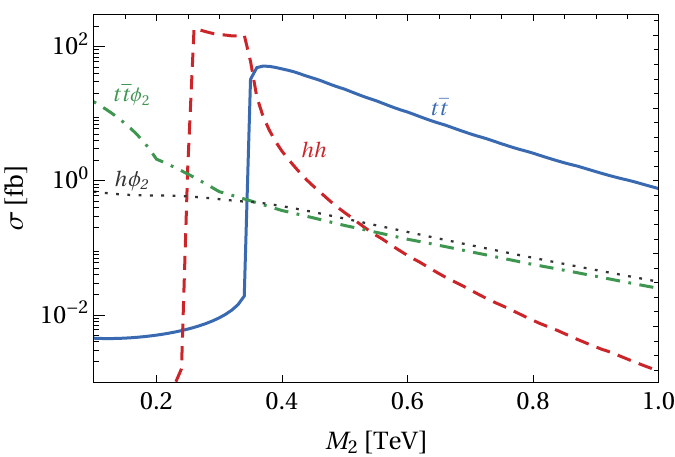}}
		\caption{Cross section of different production channels of the scalar $\phi_i$ at $\sqrt{s}=14$ TeV, evaluated for $\Lambda_1=2, \Lambda_2=1$ TeV, $c_\alpha=0.5$}
		\label{fig:xsecMS}
	\end{figure*}
	\begin{figure*}[tb]
		\centering 
		\subfloat[$hh$ ]{\includegraphics[width=0.46\textwidth]{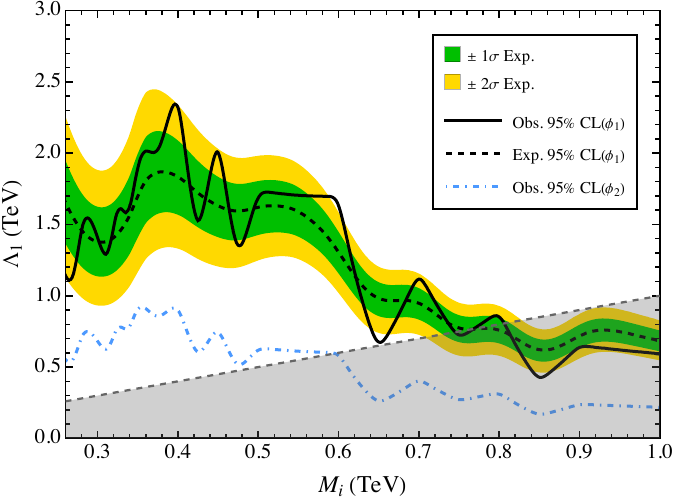}\label{fig:excluhh}}\hspace{1cm}
		\subfloat[$t\bar{t}$]{\includegraphics[width=0.46\textwidth]{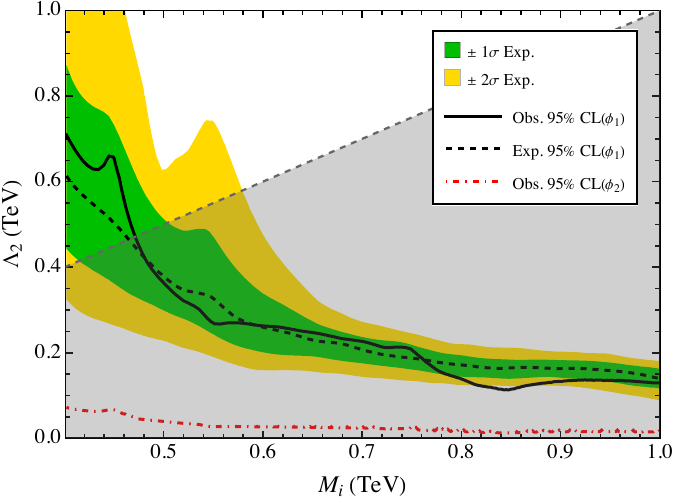}\label{fig:exclutt}}
		\caption{ Observed (solid lines) and expected (dashed lines) 95\% confidence level (CL) exclusion limits on the effective new physics scales, $\Lambda_1$ and $\Lambda_2$, as a function of the scalar mass $M_{i}$ within the 2ST framework. The green and yellow bands represent the $\pm 1\sigma$ and $\pm 2\sigma$ uncertainties on the expected limits, respectively. The limits are evaluated using benchmark parameter choices of $\cos\alpha = 0.9$ and $\Lambda_2 = 10\text{ TeV}$ for the di-Higgs ($hh$) channel, and $\cos\alpha = 0.1$ and $\Lambda_2 = 10\text{ TeV}$ for the di-top ($t\bar{t}$) channel. The gray shaded region ($\Lambda_i \le M_{i}$) indicates the kinematic domain where the effective field theory (EFT) approximation breaks down.}\label{fig:exclu}
	\end{figure*}

	Beyond the inclusive production of the new scalar, we also study the resonant production of di-Higgs and di-top final states mediated by $\phi_i$. Because the decay width of the new scalars remain narrow ($\Gamma_{i} / M_{i} \ll 1$) in the parameter space of our interest, we assume the narrow-width approximation (NWA). This allows us to factorise the signal yield into a product of the production cross section and the decay BR:
	\begin{align*}
		\sigma(pp \to \phi_i \to XX) \approx \sigma(pp \to \phi_i) \times \text{BR}(\phi_i \to XX),    
	\end{align*}
	where $XX \in \{hh, t\bar{t}\}$. Fig.~\ref{fig:xsecMS} shows these cross sections as functions of $M_{i}$. The phenomenologically interesting region is the intermediate mass window, $2m_h < M_{i} < 2m_t$. Once $M_{i}$ crosses the kinematic threshold for top-pair production ($M_{i} > 2m_t$), the $\phi_i \to t\bar{t}$ partial width grows rapidly. This rapid expansion of the total decay width causes a sharp drop in $\text{BR}(\phi_i \to hh)$ and a corresponding suppression of the di-Higgs signal, allowing the resonant $t\bar{t}$ final state to become overwhelmingly dominant. In addition to the standard di-Higgs and di-top channels, we evaluate the tri-Higgs ($h\phi_i\to hhh$) signature. This novel topology arises from the associated production mechanism ($pp \to h\phi_i$) followed by the decay, $\phi_i \to hh$. 
	
	At $\sqrt{s}=14~\mathrm{TeV}$, the scalars 
	can lead to sizeable enhancements in the di-Higgs rate. For a $300$~GeV $\phi_1$, the di-Higgs production can be about an order of magnitude larger than the SM prediction, $\sigma_{NNLO}^{\rm SM}(pp\to hh)\simeq 40~\mathrm{fb}$~\cite{deFlorian:2013jea}. For example, we find an enhancement of $\sim 12$ for the benchmark choice $\Lambda_1=2~\mathrm{TeV}$ and $\Lambda_2=1~\mathrm{TeV}$. For $\phi_2$, the enhancement is moderate, reaching a factor of about $4$ relative to the SM expectation for $M_{2}=300$~GeV.
	
	The tri-Higgs production in the SM is small, $\sigma_{\rm SM}(pp\to hhh)\simeq 0.1~\mathrm{fb}$ \cite{Abouabid:2024gms}. For $M_{1}=300~\mathrm{GeV}$, we obtain $gg\to h\phi_1\to hhh$ cross section of approximately $1.6~\mathrm{fb}$, corresponding to an enhancement by a factor of $\sim 15$ over the SM prediction. For $\phi_2$, this enhancement can reach a factor of approximately $8$.
	
	To illustrate the potential event yield at the HL-LHC, we consider the benchmark scenario in which the new scalar decays exclusively into a Higgs-boson pair, $\mathrm{BR}(\phi_i\to hh)=1$. Taking an integrated luminosity of $\mathcal{L}=3000~\mathrm{fb}^{-1}$ and requiring all three Higgs bosons to decay into $b\bar b$, with $\mathrm{BR}(h\to b\bar b)\simeq 0.58$, the expected number of signal events can be estimated at around 370 $(\mathcal{N}=\sigma_{LO}\times BR(S_0\to hh)\times BR(h\to bb)^3\times \mathcal{L}\times \epsilon)$ where $\epsilon=0.40$ denotes the assumed overall signal efficiency. Thus, even after accounting for the Higgs-boson branching fractions and a representative selection efficiency, several hundred tri-Higgs signal events could be expected at the HL-LHC for this benchmark point.
	
	\begin{figure}[t]
		\centering
		\setkeys{Gin}{width=0.25\textwidth} 
		\subfloat[$M_1=300~\text{GeV}$]{\includegraphics[width=0.48\columnwidth,height=0.47\columnwidth]{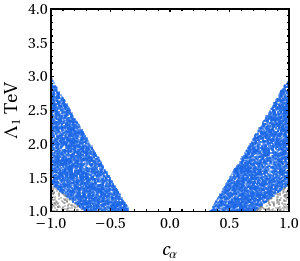}\label{fig:hlreach300a}}\hfill
		\subfloat[$M_2=300~\text{GeV}$]{\includegraphics[width=0.48\columnwidth,height=0.47\columnwidth]{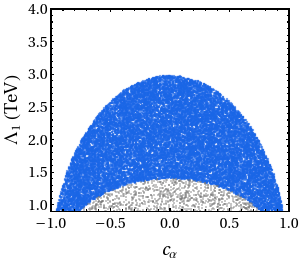}\label{fig:hlreach300b}}\\
		\subfloat[$M_1=300~\text{GeV}$]{\includegraphics[width=0.48\columnwidth,height=0.47\columnwidth]{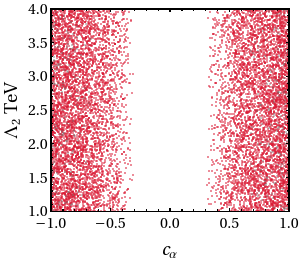}\label{fig:hlreach300c}}\hfill
		\subfloat[$M_2=300~\text{GeV}$]{\includegraphics[width=0.48\columnwidth,height=0.47\columnwidth]{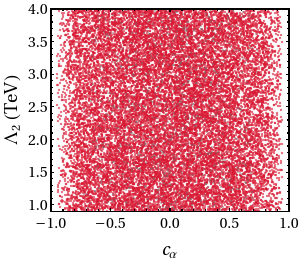}\label{fig:hlreach300d}}\\
		\subfloat[$M_1=300~\text{GeV}$ ]{\includegraphics[width=0.48\columnwidth,height=0.47\columnwidth]{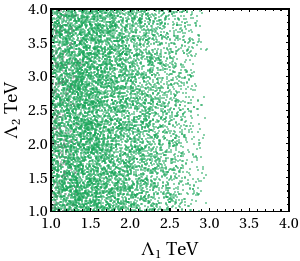}\label{fig:hlreach300e}}\hfill
		\subfloat[$M_2=300~\text{GeV}$]{\includegraphics[width=0.48\columnwidth,height=0.47\columnwidth]{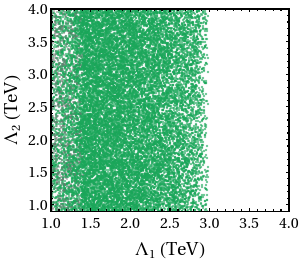}\label{fig:hlreach300f}}
		\caption{\label{fig:hllhc_2st_hh} Parameter scan showing the projected sensitivity ($\geqslant 2\sigma$) at the HL-LHC below the $t\bar{t}$ threshold ($M_{1,2}=300$ GeV) in the di-Higgs channel. The coloured points satisfy Eq.~\eqref{eq:HLLHCcondition}, except the grey points, which mark the region where the EFT breaks down. As expected, this channel is suitable for probing $\Lambda_1$ (top row), but is insensitive to $\Lambda_2$ (middle row), as confirmed by the $\Lambda_1-\Lambda_2$ scatter plots (bottom row).}
	\end{figure}
	
	\begin{figure}[t]
		\centering
		\setkeys{Gin}{width=0.25\textwidth} 
		\subfloat[$M_2=450~\text{GeV}, t\bar{t}$]{\includegraphics[width=0.48\columnwidth,height=0.47\columnwidth]{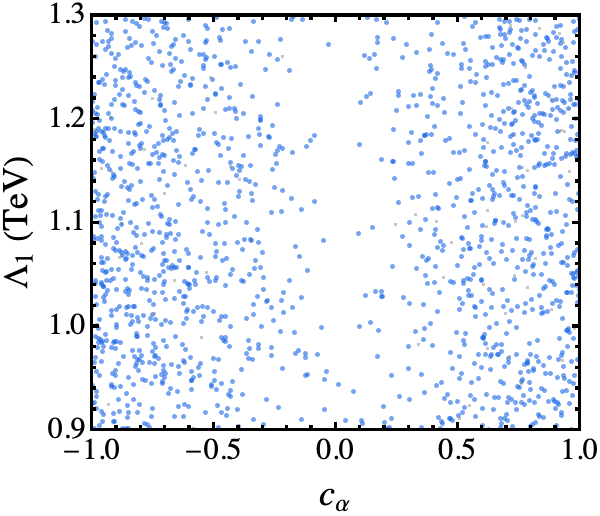}\label{fig:hlreach450a}}\hfill
		\subfloat[$M_1=450~\text{GeV}, t\bar{t}$]{\includegraphics[width=0.48\columnwidth,height=0.47\columnwidth]{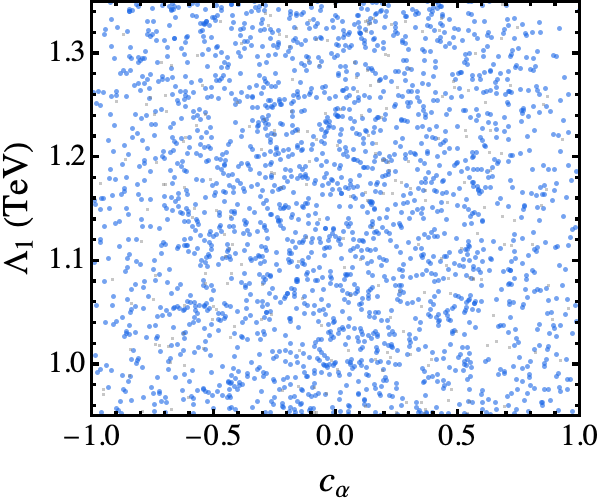}\label{fig:hlreach450b}}\\
		\subfloat[$M_2=450~\text{GeV}, t\bar{t}$]{\includegraphics[width=0.48\columnwidth,height=0.47\columnwidth]{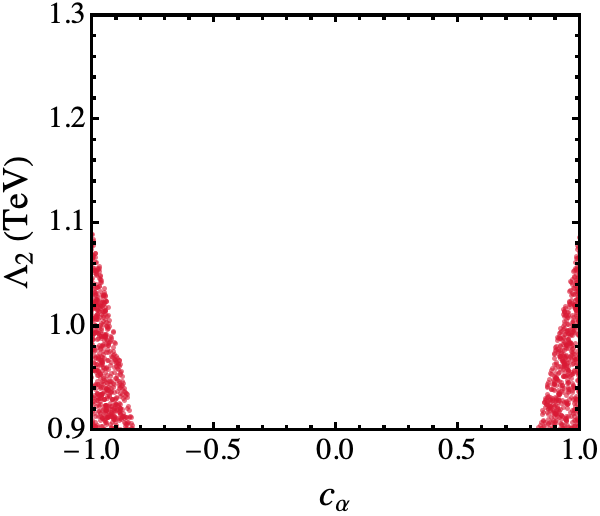}\label{fig:hlreach450c}}\hfill
		\subfloat[$M_1=450~\text{GeV}, t\bar{t}$]{\includegraphics[width=0.48\columnwidth,height=0.47\columnwidth]{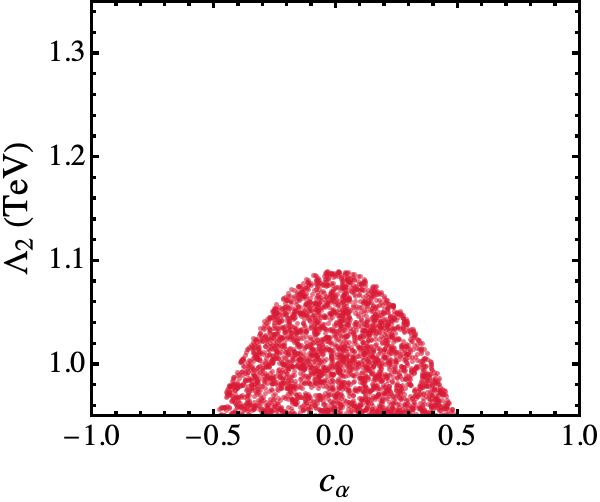}\label{fig:hlreach450d}}\\
		\subfloat[$M_2=450~\text{GeV}, t\bar{t}$]{\includegraphics[width=0.48\columnwidth,height=0.47\columnwidth]{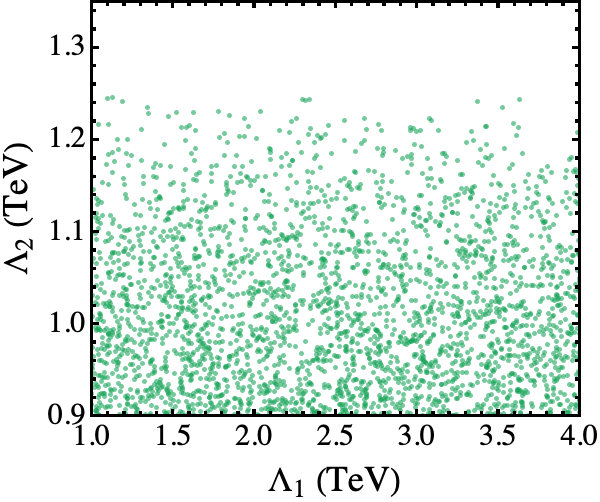}\label{fig:hlreach450e}}\hfill
		\subfloat[$M_1=450~\text{GeV}, t\bar{t}$]{\includegraphics[width=0.48\columnwidth,height=0.47\columnwidth]{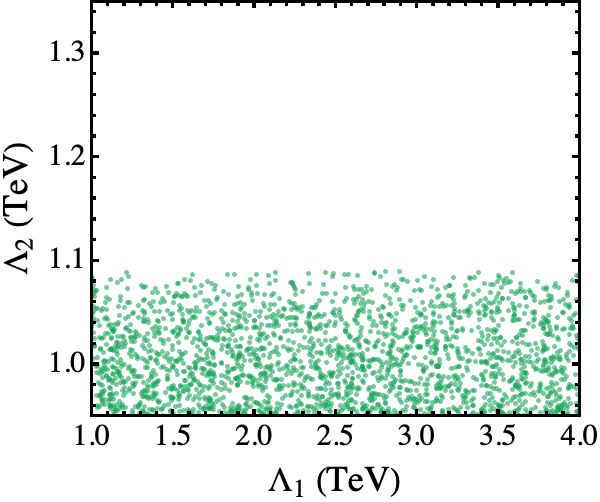}\label{fig:hlreach450f}}
		\caption{\label{fig:hllhc_2st_tt} Parameter scan showing the projected sensitivity ($\geqslant 2\sigma$) at the HL-LHC above the $t\bar{t}$ threshold ($M_{1,2}=450$ GeV) in the di-Higgs channel. The coloured points satisfy Eq.~\eqref{eq:HLLHCcondition}. This channel is insensitive to $\Lambda_1$ (top row) but can probe $\Lambda_2$ (middle row), as confirmed by the $\Lambda_1-\Lambda_2$ scatter plots (bottom row).}
	\end{figure}
	
	\section{Exclusion Limits }
	\label{sec:exclu}
	\noindent
	Once produced at the LHC, the new scalars dominantly decay via $\phi_i \to hh$ and $\phi_i \to t\bar{t}$. This lets us leverage existing experimental searches for di-Higgs and di-top resonances to constrain the 2ST parameters, specifically the cutoff scales $\Lambda_i$, as we did for the 1ST scenario in Ref.~\cite{Danielsson:2026ilg}. We recast recent 13 TeV results from the ATLAS collaboration for these specific final states:
	\begin{itemize}
		\item[---] \textit{ATLAS di-Higgs search~\cite{ATLAS:2023vdy}:} The ATLAS collaboration has searched for both resonant and non-resonant di-Higgs production in the $b\bar{b}\gamma\gamma$ final state, using $139\text{ fb}^{-1}$ of integrated luminosity. The study provides model-independent upper limits on the cross section for a new hypothetical scalar resonance in the mass range $250 \leq m_X \leq 1000$ GeV. We extract the observed and expected cross section limits, along with their associated uncertainty bands, directly from the \textsc{HEPData} database and recast them on our $\sigma(pp \to \phi_i \to hh)$ estimations.
		
		\item[---] \textit{ATLAS di-top resonance search~\cite{ATLAS:2024vxm}:} Similarly, the ATLAS collaboration has searched for heavy scalar and pseudoscalar resonances decaying into top-quark pairs, for $140\text{ fb}^{-1}$ of data. They obtain upper limits on the coupling modifier $g_{Ht\bar{t}}$ as a function of the resonance mass ($400 \leq m_H \leq 1000$ GeV) for various relative decay widths ($\Gamma/M$). Because we operate within the NWA, we systematically extract the limits corresponding to the $\Gamma/M = 1\%$ benchmark and recast them into bounds on our model parameters.
	\end{itemize}
	
	By requiring our theoretical cross sections to respect these experimental upper bounds, we derive lower limits on the cutoff scales $\Lambda_i$ as functions of the scalar mass $M_{i}$ (see Fig.~\ref{fig:exclu}). In Fig.~\ref{fig:excluhh}, we apply the $hh$ resonance limits to constrain $\Lambda_1$. For this analysis, we set the mixing parameter at $c_\alpha = 0.9$ and the second cutoff scale at $\Lambda_2 = 10$ TeV. This parameter choice is deliberate: a large $c_\alpha$ enhances $\text{BR}(\phi_1 \to hh)$, while a large $\Lambda_2$ suppresses the $\phi_1 t\bar{t}$ coupling, effectively decoupling the competing $t\bar{t}$ decay channel. Together, these conditions enhance the yield for $\sigma(pp \to \phi_1 \to hh)$. As a result, $\Lambda_1$ gets a strong lower limit across the entire mass range (shown by the solid black curve). Now, since the $\phi_2 \to hh$ coupling scales with $s_\alpha$, our choice of $c_\alpha = 0.9$ (which implies a small $s_\alpha$) naturally suppresses $\text{BR}(\phi_2 \to hh)$. With a much smaller theoretical cross section, the $\phi_2$ scalar easily evades the experimental bounds, meaning the required lower limit on $\Lambda_1$ is considerably relaxed compared to $\phi_1$ (the blue curve in Fig.~\ref{fig:excluhh}). 
	
	In Fig.~\ref{fig:exclutt}, we use to the $t\bar{t}$ resonance limits to constrain $\Lambda_2$. For that, we invert the strategy: we first choose $\Lm_1=10$ TeV. To maximise $\text{BR}(\phi_1 \to t\bar{t})$, we set $c_\alpha= 0.1$. Above $2m_t$, where the $t\bar{t}$ decay dominates, arbitrarily increasing $\Lambda_1$ yields negligible enhancements to the $t\bar{t}$ signal. However, decreasing $\Lambda_1$ would artificially enlarge the $hh$ partial width, diluting the $t\bar{t}$ branching fraction. Therefore, $\Lambda_2$ must be small if the cross section in this mode is large. Because the $\phi_2$ contribution is dictated by $s_\alpha$, its corresponding limit requires an even sharper drop, as shown by the red curve in the same plot.
	
	\section{Projected HL-LHC Sensitivity}
	\label{sec:reach}
	\noindent
	We now evaluate the discovery potential for this scenario at the High-Luminosity LHC (HL-LHC). Specifically, we identify regions of the 2ST parameter space that remain strictly consistent with current experimental data but can be probed with more than $2\sigma$ confidence level (CL) at the HL-LHC. To estimate the projected reach, we perform a straightforward luminosity scaling of the current expected limits. We assume that a parameter point will be accessible at the HL-LHC if its theoretical signal rate satisfies the following condition:
	\begin{equation}\label{eq:HLLHCcondition}
		\sqrt{\frac{L_p}{L_f}} (\sigma \times \text{BR})_\text{exp} < (\sigma \times \text{BR})_\text{th} < (\sigma \times \text{BR})_\text{obs},
	\end{equation}
	where $(\sigma \times \text{BR})_\text{obs}$ and $(\sigma \times \text{BR})_\text{exp}$ are the observed and expected upper bounds at the present integrated luminosity $L_p = 140~\text{fb}^{-1}$ respectively, and $L_f = 3000~\text{fb}^{-1}$ denotes the target HL-LHC luminosity. The theoretical prediction, $(\sigma \times \text{BR})_\text{th}$, is computed within our framework as a function of the underlying model parameters. Scaling by the square root of the luminosity ratio reduces the current expected limit by roughly 4.6, defining the wedge of parameter space that will become sensitive in the HL-LHC era. 
	
	To find the discovery potential over the range, we perform a random scan over three key parameters: the two heavy scales ($\Lambda_1$, $\Lambda_2$) and the mixing angle:
	\begin{equation}
		\Lambda_1,, \Lambda_2 \in [1~\text{TeV},, 4~\text{TeV}], \quad \cos\alpha \in [-1,, 1].
	\end{equation}
	The remaining mass parameters are fixed at two benchmarks: $M_{i} = 300$ GeV and $450$ GeV. The $300$ GeV benchmark lies in the intermediate region where di-Higgs production dominates ($m_i> 2m_h$), while the $450$ GeV benchmark probes the regime where the di-top decay channel fully opens ($m_i> 2m_t$). 
	
	We present the results of this scan as scatter plots projected onto two-parameter planes in Figs.~\ref{fig:hllhc_2st_hh} and~\ref{fig:hllhc_2st_tt}. The coloured data points correspond to parameter combinations that satisfy Eq.~(\ref{eq:HLLHCcondition}) and thus represent the regions probeable at the HL-LHC. Because $\Lambda_1$ directly enters the couplings of the new scalars to the Higgs boson, the di-Higgs channel (Fig.~\ref{fig:hllhc_2st_hh}) shows sharply defined regions in the $\Lambda_1-\cos\alpha$ plane, as the signal yield strongly depends on it. However, it is insensitive to $\Lambda_2$, which enters indirectly via the total decay widths of these scalars. Conversely, the top-pair channel can probe $\Lambda_2$ well (Fig.~\ref{fig:hllhc_2st_tt}), but is insensitive to $\Lambda_1$ for the same reason. Thus, these two channels are complementary.
	
	\section{Summary and Discussion}
	\label{sec:summary}
	\noindent In this paper, we studied dynamically varying fundamental parameters in the SM, focusing on the Higgs quartic coupling, $\lambda_h$, and the top-quark Yukawa coupling, $\lambda_t$. We assume that dimensionless couplings in a fundamental ultraviolet theory arise dynamically from the vacuum expectation values of heavy scalar fields, and use a bottom-up effective field theory to study the collider signatures at the LHC. Our previous work introduced the tightly constrained one-scalar theory (1ST), where a single scalar governs both couplings~\cite{Danielsson:2026ilg}. Here we study the two-scalar theory (2ST), which has richer phenomenology. In the 2ST framework, the variations of $\lambda_h$ and $\lambda_t$ decouple and are governed by two scalar fields that mix to form the physical mass eigenstates, $\phi_1$ and $\phi_2$. This decoupling relaxes the rigid correlations of the 1ST, yielding a flexible parameter space governed by the mixing angle $\alpha$ and the effective cutoff scales, $\Lambda_1$ and $\Lambda_2$.
	
	We evaluated the production cross sections and decay branching ratios across the 2ST parameter space. The decay channels depend sharply on kinematics. The resonant di-Higgs ($hh$) mode dominates in the intermediate mass range $2m_h<M_i<2m_t$, while the di-top ($t\bar{t}$) channel takes over once the top-pair kinematic threshold ($2m_t<M_i$) is crossed. Using recent ATLAS $hh$ and $t\bar{t}$ resonance searches, we set mass-dependent lower bounds on $\Lambda_1$ and $\Lambda_2$.
	
	Beyond standard $2 \to 2$ resonance searches, the model also predicts new kinematic signatures. The most notable is the tri-Higgs ($hhh$) final state, produced via the associated production of a heavy scalar with the SM Higgs boson ($pp \to h\phi_i$), followed by decay of the heavy scalar ($\phi_i \to hh$). These multi-Higgs signatures give background-suppressed probes of varying couplings.
	
	Our parameter scan for the High-Luminosity LHC (HL-LHC) shows why we need a multi-channel analysis strategy. The orthogonality between the Higgs-sector and top-sector cutoff scales means that neither the $hh$ nor the $t\bar{t}$ channel alone can probe the full theory. As the HL-LHC collects its expected $3000~\text{ fb}^{-1}$ of data, combined measurements of resonant $hh$ and $t\bar{t}$ channels, together with rare multi-Higgs topologies, will be needed. If SM couplings are dynamical at the TeV scale, these combined searches offer a direct window into how they arise.
	
	\section*{Acknowledgments}
	\noindent
	We acknowledge support from the Royal Society of Arts and Sciences of Uppsala (KVSU). T.M. acknowledges partial support from the SERB/ANRF, Government of India, through the Core Research Grant No.~CRG/2023/007031.
	
	\appendix
	
	\section{}\label{sec:appendix}
	\label{app:amplitude}
	
	\begin{figure*}[tb]
		\centering
		
		\tikzset{every picture/.style={scale=0.85, transform shape}}

		\subfloat[]{
			\begin{tikzpicture}[baseline=(current bounding box.center)]
				\begin{feynman}
					\vertex (g1) at (-1.2, 0.8); 
					\vertex (g2) at (-1.2, -0.8);
					\vertex (a) at (0, 0.8); 
					\vertex (b) at (1.5, 0); 
					\vertex (c) at (0, -0.8);   
					\vertex (i) at (4, 0.8); 
					\vertex (j) at (4, -0.8); 
					\vertex (h) at (3, 0);
					
					\diagram*{
						(g1) -- [gluon] (a), (g2) -- [gluon] (c),
						(a) -- [fermion] (b) -- [fermion] (c) -- [fermion] (a),
						(b) -- [scalar, edge label = $h$] (h),
						(h) -- [scalar, edge label = $\phi_i$, thick, blue] (i), 
						(h) -- [scalar, edge label' = $h$] (j),
					};
					
					\filldraw[fill=purple] (h) circle (2.5pt);
				\end{feynman}
			\end{tikzpicture}
			\label{fig:tri1hs}
		}
		\hfill 
		\subfloat[]{
			\begin{tikzpicture}[baseline=(current bounding box.center)]
				\begin{feynman}
					\vertex (g1) at (-1.2, 0.8);
					\vertex (g2) at (-1.2, -0.8);
					\vertex (a) at (0, 0.8);
					\vertex (b) at (1.5, 0); 
					\vertex (c) at (0, -0.8);   
					\vertex (h) at (3.1, .8) {$h$}; 
					\vertex [blue](k) at (3.1,-0.8) {$\phi_i$};
					
					\diagram*{
						(g1) -- [gluon] (a), (g2) -- [gluon] (c),
						(a) -- [fermion] (b) -- [fermion] (c) -- [fermion] (a),
						(b) -- [scalar] (h), 
						(b) -- [scalar, thick, blue] (k),
					};
					
					\filldraw[fill=purple] (b) circle (2.5pt);
				\end{feynman}
			\end{tikzpicture}
			\label{fig:tri2hs}
		}
		\hfill
		\subfloat[]{
			\begin{tikzpicture}[baseline=(current bounding box.center)]
				\begin{feynman}
					\vertex (g1) at (-1.2, 0.8);
					\vertex (g2) at (-1.2, -0.8);
					\vertex (a) at (0, 0.8);  
					\vertex (b) at (1.2, 0);  
					\vertex (c) at (0, -0.8); 
					\vertex (h) at (2.2, 0);
					\vertex (i) at (3.2, 0.7);
					\vertex (j) at (3.2, -0.7);
					\diagram*{
						(g1) -- [gluon] (a),
						(g2) -- [gluon] (c),
						(a) -- [fermion] (b) -- [fermion] (c) -- [fermion] (a),
						(b) -- [scalar, blue, edge label'=$\phi_i$] (h),
						(h) -- [fermion, edge label=$t$] (i),
						(h) -- [anti fermion, edge label'=$\bar{t}$] (j),
					};
					\filldraw[fill=purple] (b) circle (1.5pt);
					\filldraw[fill=purple] (h) circle (1.5pt);
				\end{feynman}
			\end{tikzpicture}
			\label{fig:tritt}
		}
		\hfill
		\subfloat[]{
			\begin{tikzpicture}[baseline=(current bounding box.center)]
				\begin{feynman}
					\vertex (g1) at (-1.2, 0.8);
					\vertex (g2) at (-1.2, -0.8);
					\vertex (a) at (0, 0.8);  
					\vertex (b) at (1.2, 0);  
					\vertex (c) at (0, -0.8); 
					\vertex (h) at (2.2, 0);
					\vertex (i) at (3.2, 0.8);
					\vertex (j) at (3.2, -0.8);
					\diagram*{
						(g1) -- [gluon] (a),
						(g2) -- [gluon] (c),
						(a) -- [fermion] (b) -- [fermion] (c) -- [fermion] (a),
						(b) -- [scalar, blue, edge label'=$\phi_i$] (h),
						(h) -- [scalar, edge label=$h$] (i),
						(h) -- [scalar, edge label'=$h$] (j),
					};
					\filldraw[fill=purple] (b) circle (1.5pt);
					\filldraw[fill=purple] (h) circle (1.5pt);
				\end{feynman}
			\end{tikzpicture}
			\label{fig:trihh}
		}\\
		\vspace{0.6cm} 
		\subfloat[]{
			\begin{tikzpicture}[baseline=(current bounding box.center)]
				\begin{feynman}
					\vertex (g1) at (-1.6, 1); \vertex (g2) at (-1.6, -1);
					\vertex (a) at (0, 1); \vertex (b1) at (1.8,1); \vertex (b2) at (1.8,-1); \vertex (c) at (0, -1);   
					\vertex (h) at (3.2, 1) {$h$}; 
					\vertex [blue](k) at (3.2,-1) {$\phi_i$};
					
					\diagram*{
						(g1) -- [gluon] (a), (g2) -- [gluon] (c),
						(a) -- [fermion] (b1) -- [fermion] (b2) -- [fermion] (c) -- [fermion] (a),
						(b1) -- [scalar] (h), 
						(b2) -- [scalar, thick, blue] (k),
					};
					
					\filldraw[fill=purple] (b2) circle (2.5pt);
				\end{feynman}
			\end{tikzpicture}
			\label{fig:box1hs}
		}
		\hfill
		\subfloat[]{
			\begin{tikzpicture}[baseline=(current bounding box.center)]
				\begin{feynman}
					\vertex (g1) at (-1.6, 1); \vertex (g2) at (-1.6, -1);
					\vertex (a) at (0, 1); \vertex (b1) at (1.8,1); \vertex (b2) at (1.8,-1); \vertex (c) at (0, -1);
					\vertex (h) at (3, 1) {$h$}; 
					\vertex [blue] (k) at (3,-1) {$\phi_i$};
					
					\diagram*{
						(g1) -- [gluon, edge label= $g$] (a),
						(g2) -- [gluon, edge label= $g$] (c),
						(a) -- [fermion, edge label= $t$] (b1) -- [fermion, edge label= $\bar{t}$] (b2) -- [fermion, edge label= $t$] (c) -- [fermion, edge label= $\bar{t}$] (a),
						(b1) -- [scalar, thick, blue] (k),
						(b2) -- [scalar] (h),
					};
					
					\filldraw[fill=purple] (b1) circle (2.5pt);
				\end{feynman}
			\end{tikzpicture}
			\label{fig:box2hs}
		}
		\hfill
		\subfloat[]{
			\begin{tikzpicture}[baseline=(current bounding box.center)]
				\begin{feynman}
					\vertex (g1) at (-1.6, 1); \vertex (g2) at (-1.6, -1);
					\vertex (a) at (0, 1); \vertex (b1) at (1.8,1); \vertex (b2) at (1.8,-1); \vertex (c) at (0, -1);
					\vertex (h) at (3, 1) {$h$}; 
					\vertex [blue] (k) at (3,-1) {$\phi_i$};
					
					\diagram*{
						(g1) -- [gluon, edge label= $g$] (a),
						(g2) -- [gluon, edge label= $g$] (c),
						(a) -- [fermion] (b1) -- [fermion] (c) -- [fermion] (b2) -- [fermion] (a),
						(b1) -- [scalar] (h),
						(b2) -- [scalar, thick, blue] (k),
					};
					
					\filldraw[fill=purple] (b2) circle (2.5pt);
				\end{feynman}
			\end{tikzpicture}
			\label{fig:box3hs}
		}
		
		\caption{$+$ all possible permutations}
		\label{fig:feynman_diagrams}
	\end{figure*}
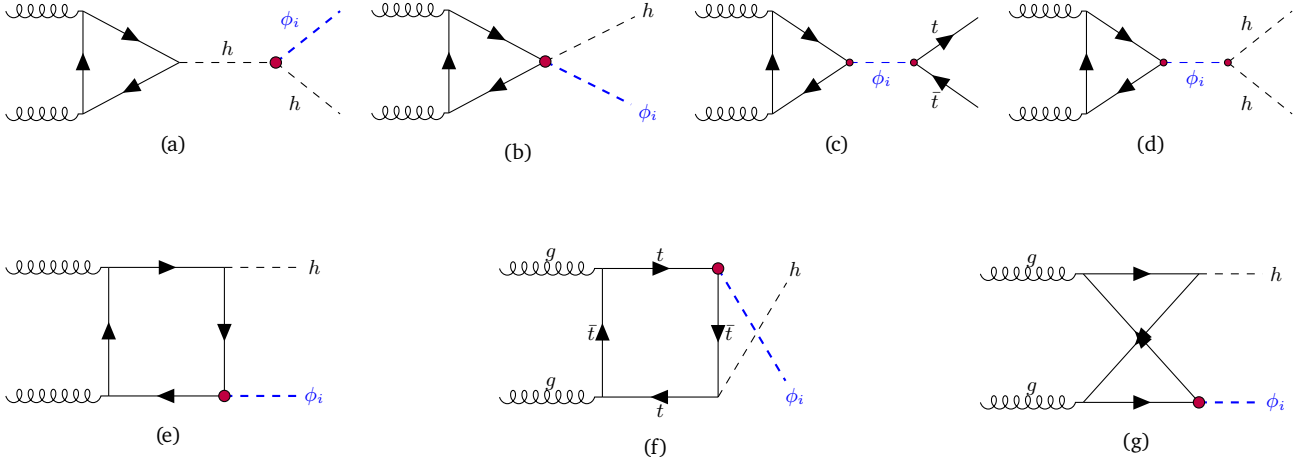
	\label{app:loopamp}
	
	Following the approach of Refs.~\cite{Glover:1987nx,Plehn:1996wb}, the partonic cross section for the gluon-fusion process $gg \to h S_0$ is governed by a scattering amplitude that, dictated by gauge invariance and Ward identities, can be decomposed into two linearly independent, Lorentz- and gauge-invariant tensor structures, $A_1^{\mu\nu}$ and $A_2^{\mu\nu}$. These tensors are constructed from the incoming gluon four-momenta, $p_a$ and $p_b$, the outgoing scalar momentum $p_c$, and the transverse momentum $p_T$ of the final state. They are explicitly given by:
	\begin{widetext}
		\begin{align}     
			A_1^{\mu\nu} = g^{\mu\nu} - \frac{p_a^\nu p_b^\mu}{(p_a \cdot p_b)};\quad\quad  
			A_2^{\mu\nu} = g^{\mu\nu} + \frac{p_c^2 p_a^\mu p_b^\nu}{p_T^2 (p_a \cdot p_b)} - \frac{2 (p_b \cdot p_c) p_a^\mu p_c^\nu}{p_T^2 (p_a \cdot p_b)} - \frac{2 (p_a \cdot p_c) p_c^\mu p_b^\nu}{p_T^2 (p_a \cdot p_b)} + \frac{2 p_c^\mu p_c^\nu}{p_T^2} \,, 
		\end{align}
	\end{widetext}
	where the transverse momentum squared can be expressed in terms of the standard partonic Mandelstam variables as $p_T^2 = (\hat{t}\hat{u} - m_h^2 M_{0}^2)/\hat{s}$. By construction, these structures satisfy the following orthogonality and normalisation conditions:
	$$A_1 \cdot A_2 = 0, \quad A_1 \cdot A_1 = A_2 \cdot A_2 = 2 \,.$$ 
	
	The transition matrix element is generated by heavy-quark loop contributions (dominated by the top quark, $m_t$) and comprises two distinct topological classes: triangle and box diagrams. In the triangle diagrams [Figs.~\ref{fig:tri1hs} and \ref{fig:tri2hs}], the incoming gluons couple to an intermediate state with a total spin projection of $S_z=0$ along the collision axis. Consequently, their contribution projects exclusively onto the $A_1^{\mu\nu}$ tensor structure. This mechanism is parameterised by a single gauge-invariant form factor, $F_\Delta$, scaled by the generalised couplings $C_\Delta^1$ and $C_\Delta^2$ corresponding to topologies (a) and (b), respectively:$$C_\Delta^1 = \frac{v}{\Lambda_0} \,, \quad \text{and} \quad C_\Delta^2 = \frac{3m_h^2}{\hat{s}-m_h^2}\left(\frac{v}{3\Lambda_0}\right) \,.$$The resulting matrix elements for the triangle contributions are:
	\begin{align}  
		\mathcal{M}^1_{\Delta} &= \frac{G_F \alpha_s}{2\sqrt{2}\pi} m_t C_\Delta^1 F_\Delta  A_1^{\mu\nu} \epsilon_\mu^a \epsilon_\nu^b \delta_{ab} \,, \\     
		\mathcal{M}^2_{\Delta} &= \frac{G_F \alpha_s}{2\sqrt{2}\pi} m_t C_\Delta^2 F_\Delta  A_1^{\mu\nu} \epsilon_\mu \epsilon_\nu \delta_{ab} \,. 
	\end{align}
	Conversely, the box diagrams [Figs.~\ref{fig:box1hs}--\ref{fig:box3hs}] exhibit a richer kinematic structure, accommodating both $S_z=0$ and $S_z=2$ gluon-gluon initial states. The box transition matrix element, therefore, spans both tensor structures and requires two gauge-invariant form factors: $F_\Box$ (associated with $A_1$) and $G_\Box$ (associated with $A_2$). Introducing the generalised box coupling $C_\Box = v / \Lambda_0$, the total box amplitude is:
	\begin{align}
		\mathcal{M}_{\Box} = \frac{G_F \alpha_s}{2\sqrt{2}\pi} \big[m_t^2 C_\Box (F_\Box A_1^{\mu\nu} + G_\Box A_2^{\mu\nu}) \big] \epsilon_\mu^a \epsilon_\nu^b \delta_{ab} \,.
	\end{align}
	Combining these kinematic structures with their respective form factors and couplings yields the fully factorised total amplitude:
	\begin{widetext}
		\begin{equation}
			\mathcal{M} = \frac{\alpha_s G_F}{2\sqrt{2}\pi} \epsilon_\mu^a \epsilon_\nu^b \delta_{ab} \bigg[ A_1^{\mu\nu} \Big( m_t (C_\Delta^1 + C_\Delta^2) F_\Delta + m_t^2 C_\Box F_\Box \Big) + A_2^{\mu\nu} m_t^2 C_\Box G_\Box \bigg]
		\end{equation}
	\end{widetext}
	The triangle and box form factors are evaluated using \textsc{Package-X}~\cite{Patel:2015tea}. They are expressed in terms of the partonic Mandelstam variables ($\hat{s}, \hat{t}, \hat{u}$) and the standard three- and four-point Passarino-Veltman scalar integrals, denoted here as $C_{ij}$ and $D_{ijk}$, respectively. They evaluate to:
	\begin{widetext}
		\begin{align}
			F_\Delta &= 2 m_t \left[2+(4 m_t^2-\hat{s}) C_{ab}\right] \,, \\
			F_\Box &= \hat{s}^{-1} \Big\{ 4\hat{s} + 8 \hat{s} m_t^2 C_{ab} - (m_h^2 + M_{0}^2 - 8 m_t^2) \big[  (m_h^2 - \hat{t}) C_{ac}+  (\hat{s} + \hat{t} - m_h^2) C_{ad} \nonumber \\
			&\quad + (m_h^2 - \hat{u}) C_{bc}+  (\hat{s} + \hat{u} - m_h^2) C_{bd}\big] \nonumber \\
			&\quad +  \big[ 16 m_t^4 \hat{s} - 2 m_t^2 \big(4 m_h^2 M_{0}^2 + 2 \hat{s}^2 + \hat{s}(\hat{t}+\hat{u}) - 4 \hat{t} \hat{u} \big) + (m_h^2 + M_{0}^2)(m_h^2 M_{0}^2 - \hat{t} \hat{u}) \big]D_{acb} \nonumber \\
			&\quad + 2 \hat{s} m_t^2 (8 m_t^2 - 2\hat{s} - \hat{t} - \hat{u}) (D_{abc} + D_{bac}) \Big\} \,, \\
			G_\Box &= (\hat{t} \hat{u} - m_h^2 M_{0}^2)^{-1} \Big\{  \hat{s} \big[ -8 m_t^2 (\hat{t}+\hat{u}) + 2 m_h^2 M_{0}^2 + \hat{t}^2 + \hat{u}^2 \big]C_{ab} \nonumber \\
			&\quad +  (m_h^2 - \hat{s} - \hat{t}) \big[ \hat{u} (\hat{u}-8 m_t^2) + m_h^2 M_{0}^2 \big]C_{ad} + (m_h^2 - \hat{s} - \hat{u}) \big[ \hat{t} (\hat{t}-8 m_t^2) + m_h^2 M_{0}^2 \big] C_{bd} \nonumber \\
			&\quad +  (8 m_t^2 - \hat{t} - \hat{u}) \big[ -2 m_h^2 M_{0}^2 + \hat{t}^2 + \hat{u}^2 \big]C_{cd} -  (m_h^2 - \hat{t}) \big[ \hat{t} (\hat{t}-8 m_t^2) + m_h^2 M_{0}^2 \big]C_{ac} \nonumber \\
			&\quad -  (m_h^2 - \hat{u}) \big[ \hat{u} (\hat{u}-8 m_t^2) + m_h^2 M_{0}^2 \big]C_{bc} + 2  m_t^2 (8 m_t^2 - \hat{t} - \hat{u}) (\hat{t} \hat{u} - m_h^2 M_{0}^2) D_{acb} \nonumber \\
			&\quad -  \Big[ 16 m_t^4 (m_h^2 M_{0}^2 - \hat{t} \hat{u}) - 2 m_t^2 \big( m_h^2 M_{0}^2 (\hat{t}+\hat{u}) - \hat{u} (-4 \hat{s} \hat{u} + \hat{t}^2 + \hat{t} \hat{u}) \big) + \hat{s} \hat{u} (m_h^2 M_{0}^2 + \hat{u}^2) \Big]D_{abc} \nonumber \\
			&\quad -  \Big[ 16 m_t^4 (m_h^2 M_{0}^2 - \hat{t} \hat{u}) - 2 m_t^2 \big( m_h^2 M_{0}^2 (\hat{t}+\hat{u}) + \hat{t} (4 \hat{s} \hat{t} - \hat{u} (\hat{t}+\hat{u})) \big) + \hat{s} \hat{t} (m_h^2 M_{0}^2 + \hat{t}^2) \Big]D_{bac} \Big\},\\
			C_{ij}
			&=
			\int d^4q
			\left[\left(q^2-m_Q^2\right)
			\left\{(q+p_i)^2-m_Q^2\right\}
			\left\{(q+p_i+p_j)^2-m_Q^2\right\}\right]^{-1},
			\\
			D_{ijk}
			&=
			\int d^4q
			\left[\left(q^2-m_Q^2\right)
			\left\{(q+p_i)^2-m_Q^2\right\}
			\left\{(q+p_i+p_j)^2-m_Q^2\right\}
			\left\{(q+p_i+p_j+p_k)^2-m_Q^2\right\}\right]^{-1}.
		\end{align}
	\end{widetext}
	
	Squaring the amplitude and averaging over initial-state colors and spins yields the differential partonic cross section:$$\frac{d\hat{\sigma}}{d\hat{t}} = \frac{G_F^2\alpha_s^2}{256(2\pi)^3}  \bigg[ \big\vert{}  (C_\Delta^1 + C_\Delta^2) \mathcal{F}_\Delta +  C_\Box \mathcal{F}_\Box \big\vert{}^2 + \big\vert{}  C_\Box \mathcal{G}_\Box \big\vert{}^2 \bigg] \,,$$where we define the scaled dimensionless form factors $\mathcal{F}_\Delta=(m_t/\hat{s})F_\Delta$, $\mathcal{F}_\Box=(m_t^2/\hat{s})F_\Box$, and $\mathcal{G}_\Box=(m_t^2/\hat{s})G_\Box$. Finally, the total partonic cross section $\hat{\sigma}(\hat{s})$ is obtained by integrating over the allowed phase space. 
	$$\begin{aligned}\hat{\sigma}(\hat{s}) = \int^{+\hat{t}_{max}}_{-\hat{t}_{max}} \frac{d\hat{\sigma}}{d\hat{t}} d\hat{t}.
	\end{aligned}$$
	The observable hadronic cross section is subsequently determined by convoluting this partonic result with the PDFs, $f(x)$, as follows
	$$\begin{aligned}
		\sigma(s) = \int_{\tau_0}^{1} d\tau \int_{\tau}^{1} \frac{dx}{x} \, f(x) \, f\!\left(\frac{\tau}{x}\right) \hat{\sigma}(\hat{s}) \,,  \end{aligned}$$
	where $\tau_0 = (m_h+M_{0})^2/s$. Assuming a proton-proton center-of-mass energy of $\sqrt{s} = 14\text{ TeV}$ alongside the mass parameters $m_h = 125\text{ GeV}$, $M_{0} = 300\text{ GeV}$, and $m_t = 172\text{ GeV}$.
	
	\bibliography{reference}
\end{document}